\documentclass[noshowpacs,prb,aps,superscriptaddress,floatfix,twocolumn,nofootinbib]{revtex4-2} 
\usepackage[utf8]{inputenc}
\usepackage[T1]{fontenc}
\usepackage{amsmath}
\usepackage{amssymb}
\usepackage{ifthen}
\usepackage{graphicx}
\usepackage{times}
\usepackage{babel}
\usepackage{color}
\usepackage{bm}
\usepackage[normalem]{ulem}
\usepackage{float}
\usepackage{placeins}
\usepackage{chngcntr}
\usepackage{enumitem}
\usepackage{multirow}
\usepackage{comment}
\usepackage{adjustbox}
\usepackage[colorlinks=true,citecolor=blue,linkcolor=black,urlcolor=blue]{hyperref}

\usepackage{blindtext}
\usepackage{float}
\usepackage{lipsum}
\usepackage{braket}

\begin{document}

	\title{Explicit proof of Anderson's orthogonality catastrophe for the one-dimensional Fermi polaron with attractive interaction}

	\author{Giuliano Orso}
	\email{giuliano.orso@u-paris.fr}
	
	\affiliation{
		Universit\'e Paris Cit\'e, Laboratoire Mat\'eriaux et Ph\'enom\`enes Quantiques (MPQ), CNRS, F-75013, Paris, France 
	}
	
	\date{\today}
	\begin{abstract}
		We provide a fully analytical derivation of Anderson's orthogonality catastrophe for the 
	    one dimensional Fermi polaron integrable model, describing a system  of $N$ spin-up  fermions,  with fixed density
		$n=N/L$,  interacting  with  a  single  spin-down  fermion via an attractive contact  potential. 
		The proof combines the determinant representations of the norm of the many-body 
		wave function and of its scalar product with the noninteracting ground state,
		obtained from the Bethe ansatz solution,  with the special properties of Cauchy matrices. 
		We derive the leading asymptotics of the two determinants in the thermodynamic limit and show that  the quasi-particle residue $Z$ 
		decays  algebraically, $Z=W N^{-\theta}$. We confirm that the Anderson exponent $\theta$ is equal to $2\delta_F^2/\pi^2$, where $\delta_F$ is the Bethe-ansatz phase shift at the Fermi edge. We also derive the asymptotic behavior of 
	   the prefactor $W$ at strong coupling  and compare it against exact numerics.
	\end{abstract}
	
	\maketitle

	\section{Introduction}
	
 The Anderson orthogonality catastrophe (AOC), first introduced by P. W. Anderson in the sixties ~\cite{Anderson_1967,Anderson_1967_2}, is a non-perturbative quantum many-body effect, describing the dramatic change in the ground state  of a system of noninteracting fermions following the scattering against a static potential. The overlap between the many-body wavefunctions before and after the perturbation vanishes in the thermodynamic limit,
	\begin{equation}\label{AOC}
		\langle \psi_\textrm{i}|\psi_\textrm{f} \rangle \sim N^{-\theta/2},
	\end{equation}
   where $N$ is the  number of fermions and $\theta$ 
	is a positive universal exponent determined by the scattering phase shift at the Fermi surface (in Eq. (\ref{AOC}) the two wavefunctions are implicitly assumed to have norm equal to one). 
	Originally inspired by the x–ray edge problem in metals \cite{Nozieres:PR1969}, the AOC   has become a cornerstone for understanding quantum quenches \cite{PhysRevX.2.041020,Schiro:PRL2014,Fogarty:PRL2020}, quantum phase transitions \cite{Zanardi:PRE2006} and quantum impurities models  \cite{affleck2009review}.

	The advent of highly controllable platforms such as ultracold atomic gases, where  interactions and confinement  can be engineered and controlled with unprecedented precision, has significantly  boost the research in impurity physics. The most studied example is the so called Fermi polaron, where a static or mobile impurity is
	coupled to a Fermi gas, which dresses it with particle-hole excitations. 
	In this case $|\psi_f\rangle$ in Eq. (\ref{AOC}) represents the ground state of the composite system at finite coupling, while $|\psi_i\rangle$ is its counterpart in the absence of coupling.  
		For recent reviews covering the experiments on polarons and the theories employed to described them, see~\cite{Massignan_2014,Schmidt_2018,Scazza_2022,Parish_2023,massignan2025RMP}.
	
    In spatial dimensions larger than one, the AOC appears
	only when the impurities has infinite mass, thus acting as a static external potential for the fermions. In the one-dimensional (1D) case, however, this effects  persists  even if the impurity has a finite mass, underlying the instability of the Fermi sea as interactions are turned on and the emergence of the Luttinger liquid scenario \cite{GiamarchiBook}. 
	
	Particularly interesting is the 1D Fermi polaron  with contact  interactions and equal masses for the impurity and the fermions. The underlying model  was shown to be exactly solvable by McGuire, who determined the energy and the effective mass of the polaron  for both repulsive~\cite{McguireJMP1965} and attractive~\cite{McGuire_1966} interactions. Later, Castella and Zotos \cite{Castella_1993} studied the spectral properties of the repulsive model,  based on the Edward's  representation
\cite{Edwards_1990} of the  ground state as a unique Slater determinant of $N$ single-particle wavefunctions. This allowed the authors of Ref. \cite{Castella_1993}  to map the 1D Fermi polaron problem back to  Anderson's original calculation for a static impurity \cite{Anderson_1967_2}, thus confirming the AOC. 

The attractive 1D Fermi polaron turned out to be  more complicated to investigate theoretically, as two of the Bethe-ansatz quasi-momenta become complex, accounting for the formation of the two-body bound state. As a consequence, the Edward's shortcut ceases to apply
and one needs to use the Takahashi's representation of the many-body wavefunction \cite{Takahashi}, as done in Ref. \cite{guanPRA2016} to explore the  correlation functions of the model. 
In a very recent article~\cite{OrsoPRA2026} we used the same  representation to extract
 the quasi-particle residue $Z$, also known as the spectral weight, which is defined as
 the normalized overlap between the ground state many-body wave function
 with its noninteracting counterpart:
 \begin{equation}
 	\label{Z}
 	Z=\frac{|\langle\psi_\textrm{NI}|\psi\rangle|^2}{\langle\psi_\textrm{NI}|\psi_\textrm{NI}\rangle \langle\psi|\psi\rangle }.  
 \end{equation}
By computing $Z$ for increasing values of $N$ and 
fitting the data tails with a power law, we provided numerical evidence that
$\ln Z\simeq -\theta \ln N$, in agreement with Eq. (\ref{AOC}).
In particular we found that the Bethe ansatz and the diagrammatic Monte Carlo method~\cite{polaron1,polaron2,Vlietinck_3D,Vlietinck_2D} yield consistent  results for $Z$, while the variational ansatz ~\cite{Chevy_2006, Combescot_2008} with one or two particle-hole excitation predicts  a finite overlap in the thermodynamic limit and  therefore fails to account for the AOC. Moreover, our results for the Anderson exponent and for the excess charge were shown to be consistent with the predictions based on boundary conformal field theory \cite{Affleck_JPA1994,AffleckProc1997}. 

	In this work we present  a fully analytical proof of the AOC for the attractive 1D Fermi polaron. The theory relies on the special properties of Cauchy matrices, as in Anderson's original derivation \cite{Anderson_1967_2}, and the Takahashi's representation of Bethe-ansatz states \cite{Takahashi}, allowing to  compute the quasi-particle residue as~\cite{OrsoPRA2026} 
	\begin{equation}
		\label{Zbis}
		Z=\frac{|\textrm{det}(V)|^2}{\textrm{det}(S)},  
	\end{equation}
	where $S$ and $V$ are square matrices of size $N$ with complex entries, coming respectively from the scalar products $ \langle\psi|\psi\rangle $ and $\langle\psi_\textrm{NI}|\psi\rangle$. 
	We show that the 
	 leading asymptotics of both determinants in the thermodynamic limit is of the form
	 \begin{equation}\label{lndetM}
	 \ln\textrm{det}(M)\simeq c_M N+d_M \ln N,
	 \end{equation}
	 where $c_M, d_M$ are coefficients that we determine analytically for each matrix $M=S,V$ (see, respectively, Eq.~(\ref{detS}) and Eq.~(\ref{finalmente}) below). Importantly, the leading coefficients are shown to be related by the condition 
	 \begin{equation} \label{cond}
	 2\Re(c_V)-c_S=0,
	 \end{equation}
	  where $\Re(\cdot)$ stands for the real part. Substituting Eq.~(\ref{lndetM}) and Eq.~
	 (\ref{cond}) in Eq.(\ref{Zbis}), we find that for  large $N$ the quasi-particle residue scales as  predicted by the AOC,  with the Anderson exponent 
	 \begin{equation}
	 \theta=d_S-2d_V.
	 \end{equation}
	 In particular, we confirm that $\theta$ depends solely on the scattering phase shift at the Fermi edge, irrespective of the presence of the two-body bound state (see Eq.~(\ref{theta}) below).

The Takahashi's representation was recently employed in Ref.~\cite{Hui:PRL2025} to compute numerically the spectral properties of the 
1D Fermi polaron in a lattice, starting from the Bethe-ansatz solution of the 1D Fermi Hubbard model. The effect of a longitudinal harmonic trap for a strongly repulsive 1D Fermi polaron was instead addressed  in Ref.~\cite{Levinsen_2015}.  

 It is worth emphasizing that in two (2D) and three (3D) spatial dimensions, a  polaron as a stable fermionic quasi-particle is formed only in the weakly interacting regime. As the impurity-fermion coupling increases, the system undergoes a first order phase transition, where the true ground state becomes a bound pair of fermions of opposite spin, dressed with particle-hole excitations of the Fermi sea. The corresponding critical value of the  interaction strength  was determined by diagrammatic Monte Carlo in 3D~\cite{polaron1,polaron2,Vlietinck_3D} and 2D~\cite{Vlietinck_2D}. 
  In contrast, for 1D models the exact Bethe-ansatz solution shows that the evolution from weak to strong coupling is a smooth crossover with no phase transition~\cite{McGuire_1966,guanPRA2016,Chang:PRA2023}. 
  The  quasi-1D regime of the Fermi polaron  has been recently investigated via the Chevy's variational ansatz (up to one particle-hole excitations of the Fermi sea), showing evidence for a polaron-to-molecule transition at strong coupling \cite{Barisic:arXiv2025}.

The article is organized as follows. In Sec. \ref{sec:model} we introduce the microscopic 
model and the explicit expressions for the norm matrix $S$ and the overlap matrix $V$
obtained from the Bethe ansatz, which are needed to extract the quasi-particle residue $Z$. Details of the derivation of the two matrices are provided in Appendix A. 
The leading asymptotics of the  $S$ and $V$ determinants in the thermodynamic limit are rigorously derived in Sec. \ref{sec:ground} and Sec. \ref{sec:overlap}, respectively. 
 In Sec. \ref{sec:Z} we collect the results to prove analytically the expected power-law decay, $Z =W N^{-\theta}$. In Sec. \ref{sec:conclusion} we provide the conclusions and an outlook.
Finally, in Appendix B we briefly recall the special properties of Cauchy matrices that are used in this work.

\section{Formulating the problem}
	\label{sec:model}
	
	In this section we briefly introduce the Hamiltonian of the system and the two Bethe-ansatz matrices, $S$ and $V$,  whose determinants allow to compute the quasi-particle residue. 
	Details of the derivation of the two matrices are given in Appendix A (see also Ref.~\cite{OrsoPRA2026} for more information).

	We consider a system of $N$ spin-up fermions in one dimension interacting with a single spin-down fermion through an attractive contact interaction. 
	The 1D model Hamiltonian is given by
	\begin{equation}
		\hat{H} =  \frac{\hat p_\downarrow^2}{2m}+  \sum_{i=1}^{N}  \frac{\hat{p}_{ i}^2}{2m}
		+ g \sum_{i=1}^{N} \delta(\hat{x}_{i}-\hat{x}_{\downarrow}) \; ,
		\label{eq:H_Gaudin}
	\end{equation}
	where $g$ is the interaction coupling constant and we assume that all fermions have equal mass $m$. This  is a special case of the Yang-Gaudin model~\cite{Yang_1967,Gaudin_1967}, which is integrable by Bethe ansatz (for a recent review see ~\cite{Guan_RMP2013}). In the following we will identify the impurity with the index value $i=N+1$, so that $\hat p_\downarrow=\hat p_{N+1}$ and $\hat{x}_{\downarrow}=\hat{x}_{N+1}$. To simplify the discussion, we will  assume that $N$ is odd, so that the total momentum $Q$  of the system in the ground state is equal to zero. 
		
	The norm matrix $S$ is given by
	\begin{equation}\label{def_S}
		S=u_{N+1} J+
		\begin{pmatrix}
			& u_1 \\
			u_2 &  \\
			& &  u_3  \\
			& &  & u_4  \\
			& &  &  &\ddots &\\
			& &   & & & u_N
		\end{pmatrix},    
	\end{equation}
	where $J$ is a matrix with all elements equal to $1$  and all omitted entries of the second matrix in the rhs of Eq.~(\ref{def_S}) are zero.  
	
Using the convention $\hbar=1$, with $\hbar$ being the reduced Planck constant,  the coefficients  $u_j$ in Eq.~(\ref{def_S}) are defined as
		\begin{equation}\label{uj}
		u_j=k_j^2+g^{\prime 2} +\frac{2 g^\prime}{L},   
	\end{equation} 
	where $g^\prime=mg/2$,  $L$ is the system size and $\{k_j\}$, with $j=1,2,\ldots,N+1$, is the set of  fermions quasi-momenta 
	obtained by solving the Bethe-ansatz equation with periodic boundary conditions:
	\begin{eqnarray}
		\frac{k_j -\lambda+i g^\prime}{k_j-\lambda-i g^\prime}=e^{i k_j L}, \; \label{BA1} \\
		\prod_{j=1}^{N+1} \frac{k_j-\lambda+i g^\prime}{k_j-\lambda-i g^\prime}=1,
		\label{BA2}
	\end{eqnarray}
	where $\lambda$ is the spin rapidity. The ground state of the system corresponds to the choice of the $N+1$ quasi-momenta  minimizing the total energy 
	\begin{equation}\label{energy} 
		E=\sum_{j=1}^{N+1} \frac{ k_j^2}{2m}.
	\end{equation}

   Since we assumed $N$ to be odd, we can set $\lambda=0$ in Eq.s (\ref{BA1}) and (\ref{BA2}), so that  if $k_j$ is a solution of Eq.~(\ref{BA1}), then $-k_j$ is also a solution with the same energy.
   As a consequence the allowed quasi-momenta can be chosen to satisfy the relation $k_{2i}=-k_{2i-1}$, which guarantees that $Q=\sum_{j=1}^{N+1} k_j =0$.
     In particular, due to the attractive nature of the interaction, the first two quasi-momenta $k_1$ and $k_2$ are purely imaginary with $k_1, k_2=\pm i g^\prime$ in the  limit $L\rightarrow \infty$. 
	These quasi-momenta give a negative
	contribution to the ground state energy (\ref{energy}), corresponding to the  energy of the two-body bound state in vacuum, $-E_B = -mg^2/4$. All the remaining quasi-momenta  $k_j$, with $j>2$, are instead real.
	
	The elements of the overlap matrix $V$ are given by
		\begin{equation}\label{V}
		V_{ij}=\frac{2g^\prime}{L} \left ( \frac{1}{k_j-q_{i+1}}- \frac{1}{k_{N+1}-q_{i+1}}\right),
	\end{equation}
	where $\{q_j\}$ is the set of allowed fermion quasi-momenta in the absence of interaction. For $g=0$  Eq. (\ref{BA1}) reduces to the condition
$e^{iq_j L}=1$, implying that $q_j=2\pi n_j/L$, with $\{n_j\}=\{0,0,1,-1, \cdots, n_N,-n_N\}$, where $n_N=(N-1)/2$. 
	
In this work we will focus on the  the thermodynamic limit, which corresponds to $N$ and $L$ both going to infinity, with the ratio $N/L$ kept fixed and equal to $k_F/\pi$, where $k_F$ is the Fermi momentum. 	
	A key role in the theoretical description of the AOC is played by the phase shifts $\{\delta_j\}$, which are defined through the relation~\cite{Castella_1993}
\begin{equation}\label{phaseshifts}
	k_j=q_j-\frac{2}{L} \delta_j.
\end{equation}

For $j\le 2$ Eq.~(\ref{phaseshifts}) yields   $\delta_j=-Lk_j/2$, implying that 
$\delta_1$ and $\delta_2$ are  purely imaginary and scale linearly with the system size, $\delta_1,\delta_2=\pm i g^\prime L/2$. 
For $j>2$, $\delta_j$ are instead real numbers, satisfying the condition $|\delta_j|\leq \pi/2$,
and can therefore be interpreted as phase shifts. To obtain their expression in the thermodynamic limit, we replace $k_j =2\pi n_j/L-2\delta_j/L$  in Eq. (\ref{BA1}), yielding
\begin{equation}\label{BAmore1}
	\frac{\pi n_j-\delta_j+i \frac{1}{\sigma}}{\pi n_j-\delta_j-i \frac{1}{\sigma}}=e^{-i2 \delta_j},
\end{equation}
where $\sigma=2/(g^\prime L)=2/(\pi \alpha N)$, with $\alpha=g^\prime/k_F$
being the dimensionless interaction parameter. In the thermodynamic limit,  the phase shifts in the lhs of  Eq. (\ref{BAmore1}) are small compared to $\sigma^{-1}$ and can therefore be neglected. Solving for $\delta_j$ we obtain
\begin{equation}\label{deltaj}
	\delta_j \simeq \frac{1}{2}i \ln\Big (\frac{\pi n_j \sigma +i}{\pi n_j \sigma-i}\Big)=-\cot^{(-1)}(\pi n_j \sigma)  \;\;\; (j>2).
\end{equation}
From Eq.~(\ref{deltaj}) one sees that  $\textrm{sign}(\delta_j)=\textrm{sign}(n_j)$. Moreover 
the absolute value of the phase shifts is close to $\pi/2$ for $n_j=\pm 1$ and diminishes as  $|n_j|$ increases.
	
	\section{Determinant of the norm matrix}
	\label{sec:ground}
	
In this section we investigate the leading large-$N$ asymptotics  of  the determinant of the matrix $S$, defined in Eq. (\ref{def_S}).  Taking into account that $u_2=u_1$, the determinant can be written explicitly as
\begin{equation}\label{detSS}
	\textrm{det}(S)=-u_1^2 \left (\prod_{j=3}^{N+1} u_j\right ) \left (\frac{2}{u_1}+\sum_{j=3}^{N+1} \frac{1}{u_j}\right ).
\end{equation}
 Replacing the sum in Eq.~(\ref{detSS})  by an integral and making use of the expression for $u_j$ in Eq.~(\ref{uj})  leads to
\begin{equation}\label{explain}
	\sum_{j=3}^N \frac{1}{u_j}\simeq L \int_{-k_F}^{k_F}\frac{dk}{2\pi}
	\frac{1}{k^2+g^{\prime 2}}
	= \frac{L}{g^\prime}\frac{1}{\pi}\arctan(\frac{k_F}{g^\prime}).
\end{equation}
Eq.~(\ref{explain}) implies that in the thermodynamic limit the sum in  Eq.~(\ref{detSS}) scales linearly with
 the system size and is therefore of the same order as $2/u_1=L/g^\prime$.  
 From Eq. (\ref{detSS}) it follows that, up to a constant term, the logarithm of $\textrm{det}(S)$ for large $N$ scales as
 \begin{equation}\label{detS0}
 \ln\textrm{det}(S)\simeq \ln\left(-4\frac{g^\prime}{L} k_F^{2(N-1)}\right)+N a,
 \end{equation}
 where
\begin{equation}\label{def_a}
	a=\frac{1}{N}\sum_{j=3}^{N+1} \ln \left(\frac{k_j^2}{k_F^2}+\frac{g^{\prime 2}}{k_F^2} \right).
\end{equation}
To estimate the coefficient $a$, we replace the quasi-momenta $\{k_j\} $ in Eq.~(\ref{def_a}) with their noninteracting counterpart $\{q_j\}$. This amounts to neglect the phase shifts in  Eq. (\ref{phaseshifts}), which is justified as their contribution only affects the next subleading term in the large-$N$ asymptotics, the constant term, which is not included in our theory. For the same reason we can replace the sum by an integral, so that 
Eq.~(\ref{def_a}) reduces to
\begin{equation}\label{res_a_1}
	a\simeq \frac{L}{2 \pi N}\int_{-k_F}^{k_F} \ln \left(\frac{q^2}{k_F^2}+\alpha^2 \right) dq.
\end{equation}	
Carrying out the integration over momentum in  Eq.~(\ref{res_a_1}) yields an analytical  expression for the $a$ coefficient
\begin{equation}\label{res_a_2} 
	a(\alpha)	=2 \alpha \arctan\left(\frac{1}{\alpha}\right)-2+\ln (1+\alpha^2),
\end{equation}
where we  have used the thermodynamic relation $N/L=k_F/\pi$. The same relation allows making the $N$-dependence in Eq.~(\ref{detS0}) more explicit 
 \begin{equation}\label{detS}
	\ln \textrm{det}(S)\simeq \ln\left( k_F^{2N}\right)+N a-\ln N,
\end{equation}
where the constant term $\ln (-4g^\prime/k_F)$ has been consistently neglected.

	\section{Determinant of the overlap matrix}
		\label{sec:overlap}
	
	In this section we derive   the large-$N$ asymptotics of  the determinant of the overlap matrix 	$V$, defined in Eq.~(\ref{V}). The calculations are significantly more involved 
	than in the previous section, so we will proceed by steps.
	Substituting Eq.~(\ref{phaseshifts}) in Eq.~(\ref{V}), we find that the entries of $V$ take the form
	\begin{equation}\label{Vbis}
	V_{ij}=\frac{g^\prime}{\pi} \left [\frac{1}{n_j- \xi_j-n_{i+1}}+ 
	\frac{1}{n_N-\xi_N+n_{i+1}} \right],
\end{equation}	
where for later convenience we have introduced the reduced phase shifts 
\begin{equation}
	\xi_j=\delta_j/\pi.
\end{equation}

From Eq.~(\ref{Vbis}) it follows that the overlap matrix can be written as 
\begin{equation}\label{defV2}
V=\frac{g^\prime}{\pi} (C+\beta \cdot w ^T),
\end{equation}
	where 
	$C$ is a square matrix of order $N$ with entries
	\begin{equation}\label{defC}
		C_{ij}=\frac{1}{n_j-\xi_j-n_{i+1}},
	\end{equation}	
	$w$ and $\beta$ are vectors,  whose $N$ components are given by, respectively,  $w_j=1$
	and 
\begin{equation}\label{beta}
	\beta_j=\frac{1}{n_N-\xi_N+n_{j+1}}.
\end{equation}

In Eq.~(\ref{defV2})  we write $C+\beta  \cdot w ^T=C(I_N+C^{-1} \beta \cdot w ^T)$, where  $C^{-1}$ is the inverse matrix  of $C$ and $I_N$ is the identity matrix of order $N$.  With the help of the identity $\textrm{det} (I_N+C^{-1} \beta \cdot w ^T)=(1+ w^T C^{-1} \beta)$, we  obtain
	\begin{equation}\label{eqdetV}
		\textrm{det}(V)=\left(\frac{g^\prime}{\pi}\right)^N  \left(1+ v \cdot \beta \right) \textrm{det}(C),
	\end{equation}
where $v$ is a vector with components $v_j=\sum_{i}^N (C^{-1})_{ij}$. Since $\xi_1^*=\xi_2$, taking the complex conjugate of  the matrix elements in  Eq. (\ref{defC}) is equivalent to exchange its first and second rows, implying that $\textrm{det}(C^*)=-\textrm{det}(C)$. Since for a general matrix $\textrm{det}(C^*)=(\textrm{det}(C))^*$, it follows that $\textrm{det}(C)$ is purely imaginary.

A key observation is that  the matrix $C$ in Eq.~(\ref{defC}) is of Cauchy type, that is it can be written as 
$C_{ij}=1/(x_i+y_j)$, where $x_i=-n_{i+1}$ and $y_j=n_j-\xi_j$.
Cauchy matrices possess remarkable properties,  which are summarized in the Appendix. 
In particular Eq.~(\ref{chauchyvj}) yields an explicit expression for the components of the vector $v$:
   \begin{equation} \label{eqvj}
   	v_j=\frac{\prod_{\ell=1}^N \left(n_\ell-\xi_\ell -n_{j+1}\right) }{\prod_{\substack{\ell=1 \\ \ell \ne j}}^N (n_{\ell+1}-n_{j+1})}.
   \end{equation}
Taking into account that $\xi_1, \xi_2$ are imaginary numbers with $\xi_2=-\xi_1$, implying that $(-\xi_1-n_{j+1})  (-\xi_2-n_{j+1})=n_{j+1}^2+|\xi_1|^2$, we find 
from Eq.~(\ref{eqvj}) that all the coefficients $v_j$ are real. 

 \subsection{Warmup}
   
   \begin{figure}
   	\includegraphics[width=\columnwidth]{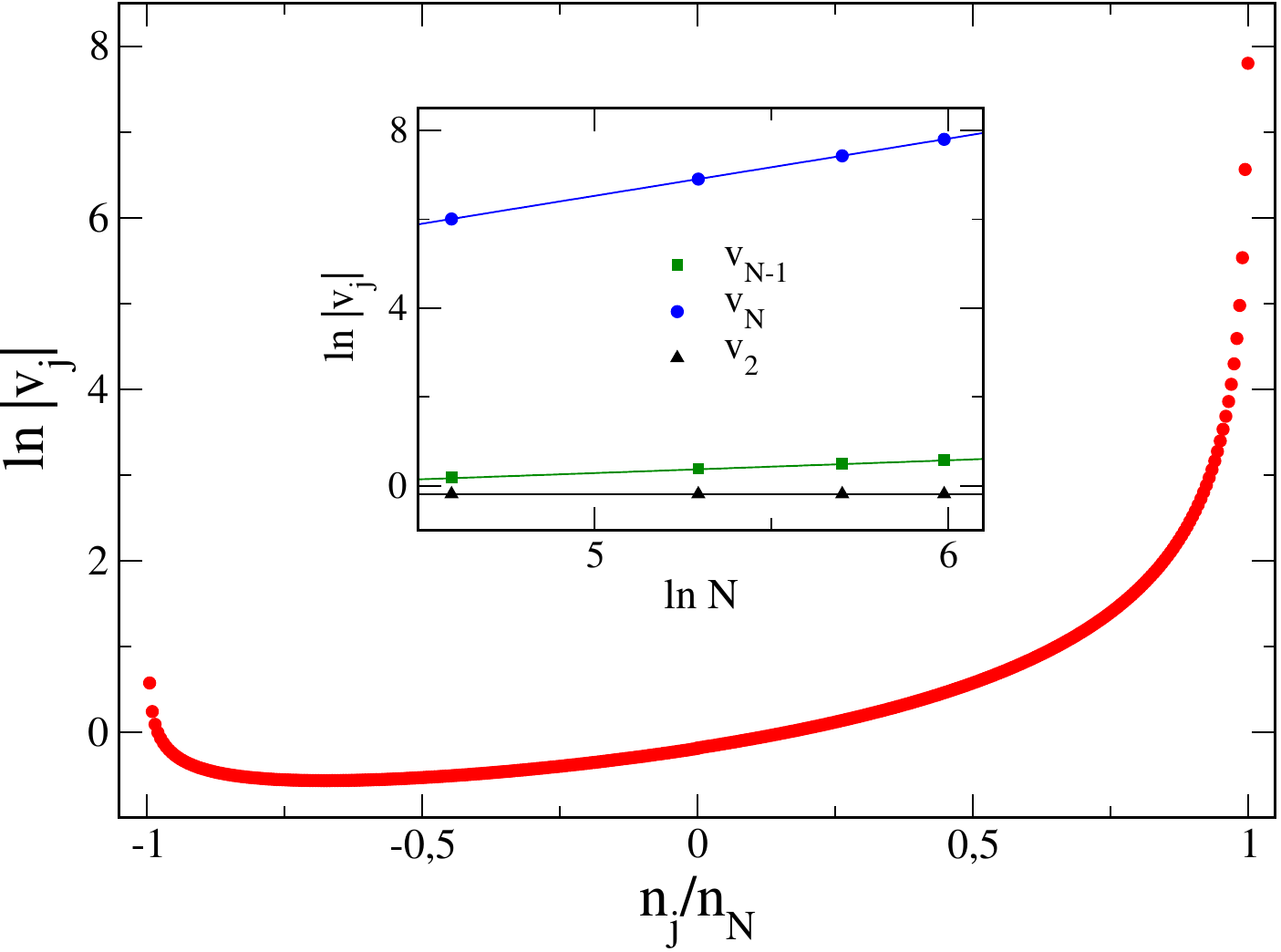}
   	\caption{Main panel: logarithm of the absolute value of the component $v_j$   (see  Eq.~(\ref{eqvj})) plotted as a function of the ratio $n_j/n_N$, for $N=399$ and $mg/(2k_F)=-4/\pi$. 
   		Inset: logarithm of the absolute value of the component $v_j$ as a function of
   		          $\ln N$ for three different values of $j=2$ (triangles), $N-1$ (squares) and $N$ (circles). The solid lines represent the linear in $\ln N$  fits of the corresponding numerical data.  
 }
   	\label{fig:scalingvj}
   \end{figure}
 
In this subsection we derive the scaling behavior of the factor $(1+ v \cdot \beta )$ in Eq.~(\ref{eqdetV}). This will give us the possibility to introduce some useful techniques
that will be used repeteadly throughout the article. 

From Eq.~(\ref{eqvj}) one can see that the dominant components of the vector $v$ come from values of  $j$ that are odd and sufficiently close to $N$.
As an example, in Fig. \ref{fig:scalingvj} we plot the logarithm of the modulus of $v_j$ as a function of the ratio $n_j/n_N$ (main panel). 
In the inset of the same figure we show that the components $v_j$ scale as a power law of the number $N$ of majority fermions, with an exponent that depends on the value of $j$. In particular $v_N$ is the component that grows faster and will be the starting point of our analysis.

Taking into account that $n_{N+1}=-n_N$, it is easy to see that all factors in the rhs of Eq.~(\ref{eqvj}) are positive for $j=N$, implying that $v_N>0$.  We find convenient to write Eq.~(\ref{eqvj}) for $j=N$  as
 \begin{equation} 
 	\ln v_N= \ln(|\xi_1|^2 +n_N^2)+\sum_{\ell=3}^{N} \ln \left (1-\frac{\xi_\ell}{n_\ell+n_{N}}\right)-\ln n_N .
 	\label{lnvn}
\end{equation}

The main contributions to the sum in  Eq.~(\ref{lnvn}) come from values of $\ell $ that are even and close to $N-1$, where the denominator takes its minimum value, $n_{N-1}+n_N=1$.  For those terms, we can approximate to leading order the reduced phase shift as $\xi_\ell \simeq -\xi_N$. Using the  expansion $\ln(1-x)\simeq -x$ for $|x|\ll 1$ in  Eq.~(\ref{lnvn}), we  
obtain
\begin{equation}\label{eqvN}
	\sum_{\ell=3}^{N} \ln \left (1-\frac{\xi_\ell}{n_\ell+n_{N}}\right) \simeq 
	\xi_N \sum_{n_\ell=-n_N+1}^{-1} \frac{1}{n_\ell+n_N}.
\end{equation} 
By setting $r=n_\ell+n_N$, the sum in the rhs of Eq.~(\ref{eqvN}) can be calculated analytically 
  \begin{equation}\label{eqvN2}
  \xi_N \sum_{r=1}^{n_N-1} \frac{1}{r}=\xi_N [\psi(n_N)-\psi(1)],
  \end{equation}
  where $\psi(z)$ is the Digamma function, which is defined as the logarithmic derivative of the Euler Gamma function, $\psi(z)=-\Gamma^\prime(z)/\Gamma(z)$. In particular for integer $z\equiv n$, the  Digamma function reduces to
  \begin{equation}
  \psi(n)=-\eta+ \sum_{r=1}^{n-1} \frac{1}{k},
  \end{equation}
  where $\eta =\psi(1)\simeq 0.577$ is the Euler-Mascheroni constant.
Substituting  Eq. (\ref{eqvN2}) in Eq. (\ref{lnvn})
 and using the asymptotic expansion $\psi(z)\simeq \ln z$, holding for $z\gg 1$, leads to
  \begin{equation}
  	\ln v_N \simeq \ln(|\xi_1|^2 +n_N^2)-\ln n_N +\xi_N \ln n_N.
  	\label{lnvn2}
  \end{equation}
  Since $|\xi_1|\sim N$, this implies that the component $v_N$ grows  with system size as
 \begin{equation}\label{scalavN}
 	\ln v_N \simeq \left (1+\xi_N\right)\ln N,
 \end{equation}
 which fully agrees with the linear in $\ln N$ fit of the numerical data shown in  
 the inset of Fig. \ref{fig:scalingvj} (solid line).
 
For $1<j<N$, it is useful to rewrite Eq.~(\ref{eqvj}) as
 \begin{equation}\label{eqlogvj}
 	\ln v_j=\ln \left[ \frac{(|\xi_1|^2 +n_{j+1}^2) \xi_{j+1}}{(-n_N -n_{j+1})n_{j+1}}\right] +\!\!\!\sum_{\substack{\ell=3 \\ \ell \ne j+1}}^{N}\frac{-\xi_\ell}{n_\ell-n_{j+1}}.
 \end{equation}
We note in passing that, since $-n_N-n_{j+1}<0$, the argument in the square bracket in Eq. (\ref{eqlogvj}) is always negative, so that $v_j<0$. It is easy to verify that this also occurs for $j=1$, implying that all components $v_j$ except the last one are negative. 
Based on Eq.~(\ref{eqlogvj}), it can be shown that $\ln |v_{N-1}| \simeq \xi_N \ln N$, more slowly than $v_N$, as displayed in the inset of Fig.\ref{fig:scalingvj} with green squares. In particular the term in the square brackets becomes independent of $N$, while the second term gives the same contribution $\xi_N \ln N$. 
In contrast, for $j\ll N$ we find that $\ln |v_j|$ saturate to a constant, with the first term in the rhs  of Eq.~(\ref{eqlogvj}) scaling as $\ln N$, while the second term yielding an opposite contribution $-\ln N$, as can be easily checked by   replacing $\xi_\ell$ with $(-1)^{\ell+1}/2$.

From Eq.~(\ref{beta}) and the above discussion, it follows that the dominant contributions to the scalar product $v\cdot \beta$ must come from values of $n_j$ close to $n_N$, where both $|v_j|$ and $|\beta_j|$ take the largest values. In this region we can approximate Eq.~(\ref{eqlogvj}) to leading order as
\begin{equation}\label{lnvjagain}
	\ln v_j\simeq \ln \left[ \frac{ n_N }{(n_j-n_N)}\right] + \sum_{\substack{n_\ell=-n_N+1 \\ n_\ell \ne - n_j}}^{-1} \frac{\xi_N}{(n_\ell+n_j)},
\end{equation}
 where we have used $|\xi_1|^2\simeq \alpha^2 n_N^2$. The sum in Eq.~(\ref{lnvjagain}) can again be evaluated
 analytically yielding
 \begin{equation}
 	\ln v_j\simeq \ln \left[ \frac{ n_N }{(n_j-n_N)}\right] + \xi_N (\psi(n_j)-\psi(n_N-n_j)),
 \end{equation}
from which we obtain
\begin{equation}\label{bv}
	\sum_{j=1}^{N-1}  v_j \beta_j \simeq - n_N e^{\xi_N \psi(n_N-1)} \sum_{n_j=1}^{n_N-1} \frac{e^{-\xi_N \psi(n_N-n_j)}}{(n_N-n_j)}.
\end{equation} 
By introducing the index $r=n_N-n_j$ and taking into account that $e^{\xi_N \psi(z)}\simeq z^{\xi_N}$ for $z\gg 1$,  Eq.~(\ref{bv}) reduces to
\begin{equation}
\sum_{j=1}^{N-1}  v_j \beta_j \simeq   - n_N n_N^{\xi_N} 	\sum_{r=1}^{n_N-1} e^{-\xi_N \psi(r)}\frac{1}{r}, 
\end{equation}  
where the sum in the rhs is convergent for infinite $n_N$, because  $e^{-\xi_N \psi(r)}\simeq r^{-\xi_N}$ at large distance. As a result we find 
\begin{equation}
	\sum_{j=1}^{N-1}  v_j \beta_j \sim N^{1+\xi_N}, 
\end{equation}
which coincides with the scaling behavior of $v_N \beta_N$ obtained from Eq.~(\ref{scalavN}), 
since $\beta_N=-1/\xi_N$ is finite for large $N$.
We therefore conclude that
\begin{equation}\label{scalingbv}
	(1+v\cdot \beta) \sim N^{1+\xi_N},
\end{equation}
 which is the key result of this subsection.
 
 \subsection{Reshuffling}
Extracting the scaling behavior of the determinant of the matrix $C$ in Eq.~(\ref{eqdetV})  is the most challenging part of the work. In this subsection we explain our  strategy, which amount to split the determinant in several factors, each of which will be studied independently in the next subsection.

We express the logarithm of the determinant of the Cauchy matrix $C$, defined in Eq. (\ref{defC}), with the help of Eq.~(\ref{cauchydet}) :
\begin{eqnarray}\label{logdet1}
	\ln \textrm{det}(C) &=& \sum_{j=2}^N\sum_{i=1}^{j-1}\left [ \ln (n_{j+1}-n_{i+1})+
	\ln (n_i-\xi_i-n_j+\xi_j) \right]\nonumber\\
	&- &\sum_{i,j=1}^N \ln \left (n_j-\xi_j-n_{i+1}\right).
\end{eqnarray}
At first look one could  expect that  the rhs Eq.~(\ref{logdet1}) is proportional to $N^2$ for large $N$, due to the presence of double sums, but sever cancellation effects occur between the first and the second rows.
  It is therefore convenient to reorganize the different contributions in  Eq. (\ref{logdet1}) to account for this fact.

We start by splitting the sums in the rhs of Eq.~(\ref{logdet1}) between indices corresponding to complex and to real phase shifts
\begin{eqnarray}
	\ln \textrm{det}(C) &=& \ln  (-\xi_1+\xi_2) +\sum_{j=2}^N\sum_{i=1}^{j-1} \ln(n_{j+1}-n_{i+1} ) \nonumber\\
	&+& \sum_{j=3}^N \left (\sum_{i=1}^{2}+\sum_{i=3}^{j-1}\right) \ln  (n_i-\xi_i-n_j+\xi_j ) \nonumber \\
	&-&\sum_{i=1}^{N} \left(\sum_{j=1}^2+\sum_{j=3}^N \right)  \ln  (n_j-\xi_j-n_{i+1}), \label{logdet2}
\end{eqnarray}
  where we made use of the fact that $n_1=n_2=0$. 
  Next, we rewrite the first double sum  
  $B\equiv \sum_{j=2}^N\sum_{i=1}^{j-1} \ln(n_{j+1}-n_{i+1} )$  in the rhs of Eq.~(\ref{logdet1}) as
  \begin{equation}\label{eqB1}
  	B\equiv  \left (\sum_{j=2}^N \sum_{\substack{i=1  \\ i \ne j}}^N-\sum_{i=3}^{N}\sum_{j=2}^{i-1}\right ) \ln(n_{j+1}-n_{i+1} ),
  \end{equation}
  and shift  the sums indices  to obtain
  \begin{eqnarray}\label{eqB2}
  	B&=&\sum_{j=3}^N \sum_{\substack{i=1  \\ i \ne j}}^N \ln(n_{j}-n_{i+1}) + \sum_{i=1}^{N-1} \ln(n_{N+1}-n_{i+1}) \nonumber \\
  	&-&\sum_{i=4}^N\sum_{j=3}^{i-1} \ln(n_{j}-n_{i}) -\sum_{j=3}^N \ln(n_{j}-n_{N+1}).
  \end{eqnarray}

  Finally, the first contribution in the last row of Eq.~(\ref{logdet1}) is recast as
  \begin{eqnarray} \label{defA}
  &-&\sum_{i=1}^{N} \sum_{j=1}^2 \ln \left (n_j-\xi_j-n_{i+1}\right)=-\sum_{i=3}^{N} \sum_{j=1}^2 \ln \left (-\xi_j-n_{i}\right) \nonumber \\
  &\;& \;\;\;-\sum_{j=1}^2 \ln  (-\xi_j) -\sum_{j=1}^2 \ln \left (-\xi_j-n_{N+1}\right). 
\end{eqnarray}
  
Inserting Eq.s (\ref{eqB2}) and (\ref{defA})  back in Eq.~(\ref{logdet2})  allows writing  
$\ln \textrm{det}(C) = \sum_{i=1}^4F_{i}$, where the quantities $F_i$ are defined as
\begin{eqnarray}\label{central}
&F_1=& \sum_{j=3}^N \sum_{i=1}^2  [\ln (-\xi_i-n_j+\xi_j ) 
-\ln (-\xi_i-n_j )] \nonumber \\
&F_2=& \sum_{j=4}^N \sum_{i=3}^{j-1} [\ln (n_i-\xi_i-n_j+\xi_j) -\ln (n_i-n_j)] \nonumber \\
&F_3=&\sum_{j=3}^N \sum_{\substack{i=1  \\ i \ne j-1}}^N [\ln(n_j-n_{i+1})-\ln(n_j-\xi_j-n_{i+1})] \\
&F_4=&-\sum_{j=3}^N \ln(-\xi_j)-\sum_{j=1}^2 \ln  (-\xi_j)  -\sum_{j=1}^2 \ln (-\xi_j+n_{N})\nonumber \\
&&+\ln (-\xi_1+\xi_2)-\sum_{j=3}^{N}  \ln(n_{j}+n_{N})+\sum_{j=2}^N \ln(-n_N-n_{j}). \nonumber
\end{eqnarray}
Since $|\xi_j|<1/2$  for $j>2$, the two contributions  in the summands of 
$F_2$ and $F_3$ tend to cancel each other when their argument becomes large (in absolute value), implying that the $N^2$ term is absent in their large-$N$ asymptotics.

\subsection{Leading asymptotics of $F_i$}

In this subsection we extract the  behavior of $\ln \textrm{det}(C)$ in the thermodynamic limit, which is at the core of our proof.  We will show that $F_1$  converges to a constant and is therefore irrelevant for our large-$N$ asymptotics, while the other contributions  $F_j$ scale  as 
\begin{eqnarray}
	&F_2 \simeq  & a_2 N +\frac{1}{2} \ln N \label{keyresultF2} \\ 
	&F_3 \simeq  & a_3 N -(\xi_N^2+\xi_N) \ln N  \label{keyresultF3}\\
	&F_4 \simeq & a_4 N -2 \ln N +i\pi n_N+ i\frac{\pi}{2} \label{keyresultF4}
\end{eqnarray}
where the coefficients $a_i$ are real functions of the interaction parameter $\alpha$  to be determined. 

For what follows, it is important to notice that the difference between the two logarithms in the summands of the functions $F_1, F_2$ and $F_3$ can be written under the form $\ln (1-x)$, whose infinite series representation is given by
\begin{equation}\label{logexpansion}
	\ln (1-x)=-\sum_{s=1}^{+\infty} \frac{x^s}{s}.
\end{equation}

We first show that the quantity $F_1$ in Eq. (\ref{central}) is  convergent in the thermodynamic limit.
Setting $x=\xi_j/(\xi_i+n_j)$ and taking into account that, for  $i=1,2$, $\xi_i$ are purely imaginary with  $|\xi_i|\sim N$, while $|\xi_j|\le 1/2$ for $j>2$, yields $|x|\leq 1/|\xi_i|\sim 1/N$. Therefore only the  first order term in Eq. (\ref{logexpansion}) contributes significantly to $F_1$. Since the absolute value of such term is upper bounded by 
\begin{equation}\label{F1bis}
	\sum_{j=3}^N \sum_{i=1}^2 \left |\frac{\xi_j}{(\xi_i+n_j)}\right |
	 \leq \sum_{i=1}^2 \frac{1}{|\xi_i|}   N=2 \frac{N}{|\xi_1|},
\end{equation}
with the rhs remaining finite in the thermodynamic limit, we conclude that $F_1$
saturates to a constant value.

We now turn to the scaling behavior of $F_2$. Making use of Eq. (\ref{logexpansion}) with  $x=(\xi_i-\xi_j)/(n_i-n_j)$ for $F_2$ in Eq. (\ref{central}), we expressed it as
\begin{equation}\label{newF2}
	F_2= -\sum_{j=3}^N \sum_{i=3}^{j-1} \sum_{s=1}^{+\infty} \frac{1}{s} \frac{(\xi_i-\xi_j)^s}{(n_i-n_j)^s}.
\end{equation}
The leading term in the large$-N$ expansion for $F_2$  can be obtained by taking the continuum limit of Eq. (\ref{newF2}). To this purpose we set $n_i=x n_N$, $n_j=y n_N$, and replace the reduced phase shifts with the thermodynamic limit result based on Eq. (\ref{deltaj}), that is  $\xi_j=\xi^\textrm{th}(n_j/n_N)$, where
\begin{equation}\label{xi}
	\xi^\textrm{th}(x)=-\frac{1}{\pi}\arctan\left (\frac{\alpha}{x}\right).
\end{equation}
We notice that only the  $s=1$ term in Eq.~(\ref{newF2}) gives a finite contribution to $a_2$ in Eq.~(\ref{keyresultF2}). For such term the integration over the $x$ variable takes the form 
 \begin{equation}\label{deff}
f(y)=-\int_{-|y|}^{|y|}dx \left(\frac{\xi^\textrm{th}(x)-\xi^\textrm{th}(y)}{x-y}\right),
\end{equation}
where the domain of integration reflects  the condition $i<j$.
Taking into account that   $n_N\simeq N/2$ for $N\gg 1$, we find
\begin{equation}\label{a2new}
	a_2=\frac{1}{2}\int_{-1}^{1}f(y)dy= -\int_{0}^{1}dy \int_{-y}^{y}dx \left(\frac{\xi^\textrm{th}(x)-\xi^\textrm{th}(y)}{x-y}\right),
\end{equation}
where in the second equality we have used the fact that the function  $f(y)$ is pair. 
The above expression for $a_2$ still involves a double integral but one of the two integrations can be done analytically, after introducing  a imaginary term $i\epsilon$ in the denominator to avoid the apparent singularity at $x=y$.
 For the first term in the integrand of Eq.~(\ref{a2new}), this requires exchanging the order of integration
 leading to
\begin{eqnarray}\label{a2w1}
	&-& \left (\int_0^1 dx\int_x^1 dy+\int_{-1}^0 dx \int_{-x}^1 dy\right) \frac{\xi^\textrm{th}(x)}{x-y+i\epsilon}\nonumber \\
&=&\int_{-1}^1 \xi^\textrm{th}(x) \ln(x-1)dx  -\ln(i \epsilon)\int _0^1  \xi^\textrm{th}(x) dx \\
&-&\int_{-1}^0  \xi^\textrm{th}(x) \ln(2x) dx\nonumber.
\end{eqnarray}
For the second contribution in Eq.~(\ref{a2new}), a direct integration over $y$ yields
\begin{equation}\label{a2w2}
	\int_{0}^{1}dy \int_{-y}^{y}dx \frac{ \xi^\textrm{th}(y) }{x-y+i \epsilon}=\int_0^1 dy\xi^\textrm{th}(y) [\ln(i \epsilon)-\ln(-2y)],
\end{equation}
which cancels  with the second and the third contributions in Eq. (\ref{a2w1}). This leads to 
\begin{equation}\label{a2ok}
	a_2=\int_{-1}^{1} \xi^\textrm{th}(x)\ln (1-x) dx =\int_0^1 \xi^\textrm{th}(x) \ln\left(\frac{1-x}{1+x}\right) dx,
\end{equation}
where in the second equality we have used the odd parity of the phase shifts in Eq.~(\ref{xi}). Notice  that the above result is independent of $\epsilon$, as the singularity in Eq.~(\ref{a2new})  is only apparent.  

\begin{figure}
	\includegraphics[width=\columnwidth]{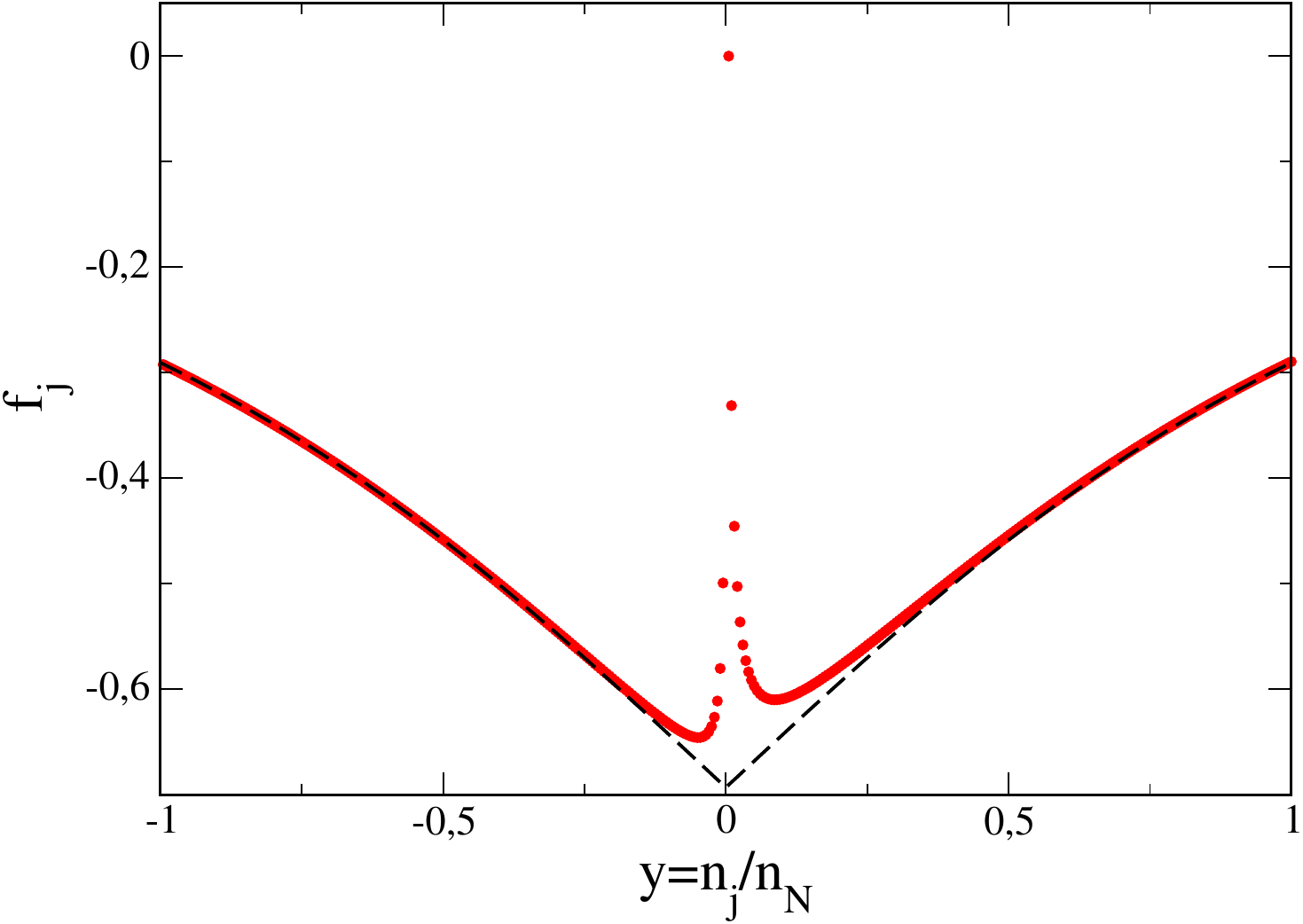}
	\caption{$j$-distribution obtained from the first order term $(s=1)$ in the expansion of $F_2$, see Eq.~(\ref{newF2}),  after summing over the index $i$. The dashed line corresponds to the continuum result, Eq.~(\ref{deff}). Data for $N=399$ and  $mg/(2k_F)=-4/\pi$.}
	\label{fig:diff2}
\end{figure}

In order to obtain the subleading logarithmic correction, we write $F_2=\sum_{j=4}^N f_j$, where $f_j$ results from the summation over the index $i$ for a fixed value of $j$.  Since $i<j$, for small values of $j$ this sum is made of only few terms, therefore replacing it by an integral introduces an error. 
To see this, we have calculated $f_j$ numerically from Eq.~(\ref{newF2}), by restricting the sum  to $s=1$. The result is displayed in   Fig.\ref{fig:diff2} as a function of $y=n_j/n_N$, showing clear deviations for $|y|\ll 1$ with respect to the 
continuum limit prediction in Eq.~(\ref{deff}). 

To proceed further, we write the deviation as $F_2-a_2 N= X+Y+...$, where  $X$ and $Y$ are the contributions obtained from  Eq.~(\ref{newF2}) for, respectively, $s=1$ and $s=2$ (we will see later that higher values of $s$ do not produce  logarithmic corrections).
For small values of $j$, the reduced phase shift can be approximated as $\xi_j \simeq  (-1)^{j+1}/2$, depending only on the sign of $j$.  As a consequence, the only finite contributions to $f_j$ come from configurations, where $n_i$ and  $n_j$
have opposite signs. Taking into account that $f(0)=-\ln 2$, we find 
\begin{equation}
X=\!\!\!	\sum_{n_j=-n_N+1}^{-1}\!\!\Big(\ln 2 +\sum_{n_i=1}^{-n_j} -x_i\Big) + \sum_{n_j=2}^{n_N-1}\Big(\ln 2+ \!\!\sum_{n_i=-n_j+1}^{-1} -x_i\Big),
\end{equation}
with $x_i=\textrm{sign}(n_i)/(n_i-n_j)$. The summation over the index $i$ can be done analytically 
\begin{eqnarray}\label{EqX}
X&=&\sum_{n_j=-n_N+1}^{-1} \Big(\ln 2- \psi(1-2n_j)+\psi(1-n_j)\Big ) \nonumber \\
&+&\sum_{n_j=2}^{n_N-1}\Big(\ln 2+ \psi(1+n_j)-\psi(2n_j)\Big).
\end{eqnarray}
To perform the residual sum over $j$ in Eq.~(\ref{EqX}), we replace the
summands by their asymptotic expansion at large $n_j$,  making use of the formula 
\begin{equation} \label{asiPg}
	\psi(z)\simeq  \ln z-\frac{1}{2z}-\frac{1}{12 z^2}, 
\end{equation}
holding for $z\gg 1$. Retaining only the leading non vanishing term yields
\begin{equation}
	X\simeq \sum_{n_j=-n_N+1}^{-1} -\frac{1}{4}\frac{1}{n_j}+\sum_{n_j=2}^{n_N-1} \frac{3}{4}\frac{1}{n_j}\simeq \sum_{n_j=2}^{n_N} \frac{1}{n_j},
\end{equation}
implying that $X\simeq  \ln N$.

The correction from $s=2$ is given by 
\begin{equation}\label{Ysimple}
	Y=	\sum_{n_j=-n_N+1}^{-1} \sum_{n_i=1}^{-n_j} -\frac{1}{2}x_i^2 + 
	\sum_{n_j=2}^{n_N-1} \sum_{n_i=-n_j+1}^{-1} -\frac{1}{2}x_i^2,
\end{equation}
where the summation over $n_i$ can again be performed analytically leading to
\begin{eqnarray}
	Y&=&\sum_{n_j=-n_N+1}^{-1} -\frac{1}{2}\Big(- \psi^\prime (1-2n_j)+\psi^\prime(1-n_j)\Big ) \nonumber \\
	&+&\sum_{n_j=2}^{n_N-1} -\frac{1}{2} \Big( \psi^\prime(1+n_j)-\psi^\prime(2n_j)\Big),
\end{eqnarray}
where $\psi^\prime(z)$ is the derivative of $\psi(z)$. 
Taking into account that $\psi^\prime(z)\sim 1/z$ for $z\gg 1$, we obtain
\begin{equation}
	Y= \sum_{n_j=-n_N+1}^{-1} \frac{1}{4}\frac{1}{n_j} +\sum_{n_j=2}^{n_N-1} -\frac{1}{4}\frac{1}{n_j} \simeq \sum_{n_j=2}^{n_N} -\frac{1}{2}\frac{1}{n_j},
\end{equation}
showing that $Y\simeq (-1/2) \ln N$ and therefore $X+Y\simeq (1/2) \ln N$. 

To recover Eq.~(\ref{keyresultF2}), we need to show that the contributions from  $s> 2$  
 in Eq.~(\ref{newF2}) do not produce additional logarithmic divergences. For a given $s$
 the contribution reads
\begin{equation}
\sum_{n_j=-n_N+1}^{-1} \sum_{n_i=1}^{-n_j} -\frac{1}{s}x_i^s + 
\sum_{n_j=2}^{n_N-1} \sum_{n_i=-n_j+1}^{-1} -\frac{1}{s} x_i^s 
\end{equation}
The summation over $n_i$ can be performed analytically, yielding
\begin{eqnarray}\label{check0}
	&& \sum_{n_j=-n_N+1}^{-1}  \frac{1}{s}\left(\zeta_H(s,1-2n_j)-\zeta_H(s,1-n_j)\right)\nonumber\\
	&+& \sum_{n_j=2}^{n_N-1} \frac{1}{s} \left( \zeta_H(s,2n_j)-\zeta_H(s,1+n_j)\right),
\end{eqnarray}
where  $\zeta_H(s,x)$ is the Hurwitz Zeta function.
Based on its asymptotics at large distance
\begin{equation}
\zeta_H(s,x)\simeq \frac{x^{1-s}}{s-1}\;\;\;\; (|x|\gg 1),
\end{equation}
 one  immediately see that the absolute values of the addends in the two sums in Eq.~(\ref{check0})  decay as $|n_j|^{1-s}$, implying that for $s>2$
 the corresponding series for $n_N\rightarrow +\infty$ are (absolutely) convergent. Hence  no additional logarithmic singularities in the $F_2$ expansion appear from cubic and higher order terms.

We next focus on the $F_3$ term. Using Eq.~(\ref{logexpansion}) with $x_i=\xi_j/(n_j-n_{i+1})$ for $F_3$ in Eq.~(\ref{central}) yields
\begin{equation}\label{F_3expa}
	F_3=\sum_{s=1}^{+\infty} \sum_{j=3}^N\frac{1}{s}\xi_j^s \sum_{\substack{i=1  \\ i \ne j-1}}^N \frac{1}{(n_j-n_{i+1})^s}.
\end{equation} 
If we try to evaluate the contribution of order $s$ in Eq.~(\ref{F_3expa}) by replacing the sums in Eq.~(\ref{F_3expa}) by integrals, as previously done for $F_2$, we obtain 
\begin{equation}\label{F_3gen}
	n_N^2 \frac{1}{n_N^s} \int_{-1}^{1} dy \frac{1}{s} (\xi^\textrm{th} (y))^s P \int_{-1}^{1}dx \frac{1}{(y-x)^s},
\end{equation}
where $P$ stands for the principal value, accounting for the fact that $i \neq j-1$. From Eq. (\ref{F_3gen})  one could naively conclude that only $s=1$ gives a finite contribution to the coefficient $a_3$, but this is untrue, because for $s>1$   a
nonintegrable singularity appears for $x=y$.  In contrast, we notice that for $s>1$ the summation over the index $i$ in Eq.~(\ref{F_3expa}) is convergent as $n_N$ tends to infinity, implying that we can extend the sum to infinity from both sides, yielding
\begin{equation}\label{series}
	\sum_{\substack{n_i=-n_N+1  \\ n_i \ne n_j}}^{n_N} \frac{1}{(n_j-n_i)^s}\simeq \sum_{\substack{n_i=-\infty  \\ n_i \ne n_j}}^{+\infty} \frac{1}{(n_j-n_i)^s}.
\end{equation} 
For $s$ even, the sum of the series in the rhs of Eq.~(\ref{series}) is equal to $2 \zeta(s)$, where $\zeta(z)$ is the Riemann Zeta function, while for $s$ odd the sum vanishes. 

Putting together the above results, we find that the coefficient $a_3$ is given by
\begin{equation}\label{a3finally}
	a_3=\frac{1}{2} \int_{-1}^{1} dy \xi^\textrm{th}(y) \ln\Big(\frac{1+y}{1-y}\Big)+ \int_{-1}^{1} dy  \sum_{k=1}^{+\infty} \frac{\zeta(2k)}{2k} 
	(\xi^\textrm{th}(y))^{2k}.
\end{equation}
The above formula can be further simplified by noting  that the sum of the series in Eq.~(\ref{a3finally}) can be evaluated analytically
\begin{equation}
	\sum_{k=1}^{+\infty} \frac{\zeta(2k)}{2k} 
	x^{2k}=\frac{1}{2} \ln\left(\frac{\pi x }{\sin (\pi x)}\right),
\end{equation}
so that we finally obtain
\begin{equation}\label{a3ok}
	a_3= \int_{0}^{1} dy \xi^\textrm{th}(y) \ln\Big(\frac{1+y}{1-y}\Big)+ \int_{0}^{1} dy   \ln\left(\frac{\pi \xi^\textrm{th}(y) }{\sin (\pi \xi^\textrm{th}(y))}\right).
\end{equation}

\begin{figure}
	\includegraphics[width=\columnwidth]{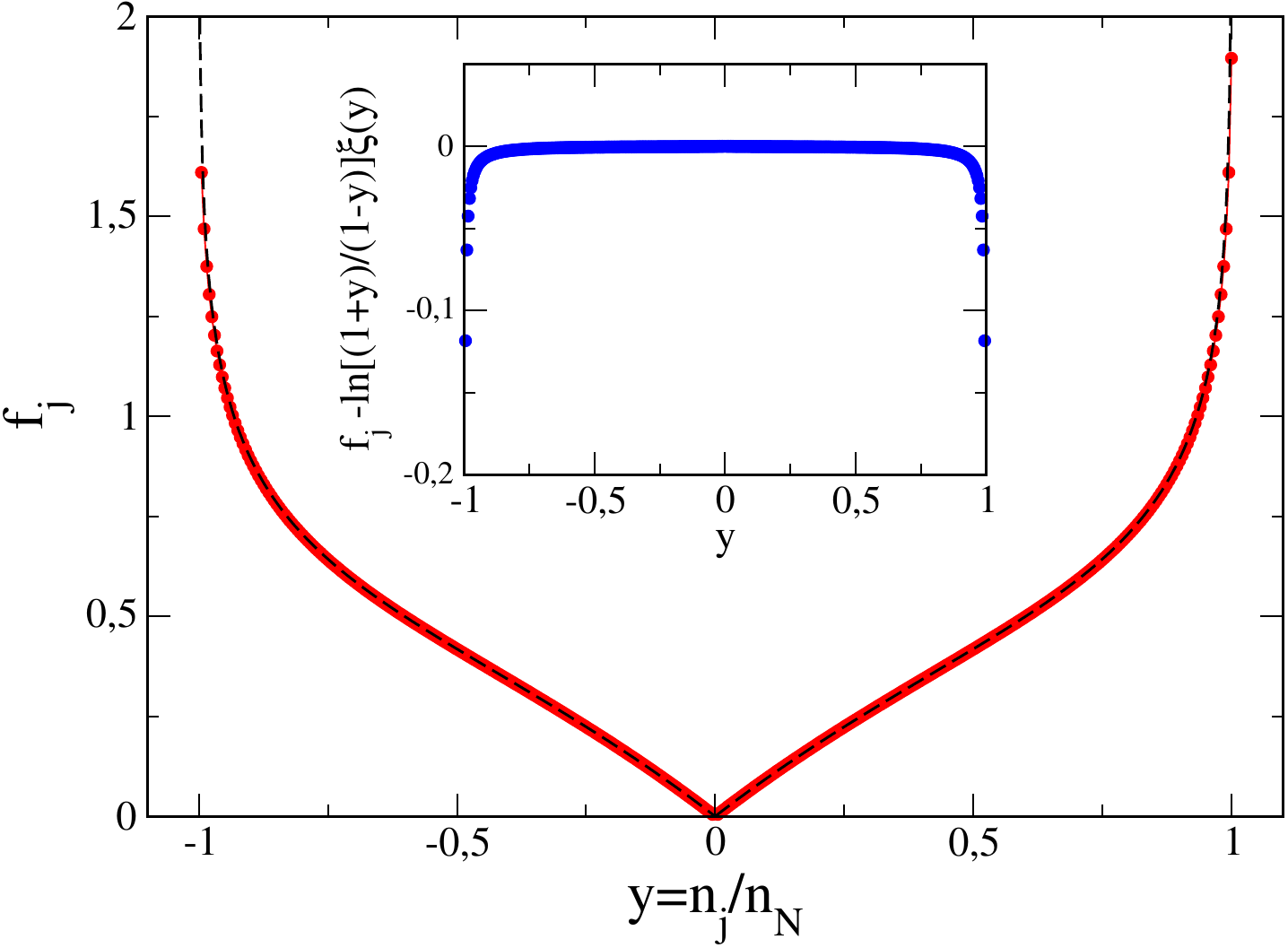}
	\caption{Main panel: $j$-distribution obtained from the first order term $s=1$ of $F_3$, see  Eq. (\ref{F_3expa}),		after summing over the index $i$, plotted as a function of $y=n_j/n_N$. The dashed line corresponds to the function $\xi^\textrm{th}(y) \ln[(1+y)/(1-y)]$ resulting from the continuum limit. Inset:  difference between the two curves.  Notice that the difference is always negative, leading to a logarithmic correction with a negative prefactor. Data for $N=399$ and $mg/(2k_F)=-4/\pi$. }
	\label{fig:diff3}
\end{figure}

Let us now evaluate the  logarithmic terms in $F_3$. We write again $F_3=\sum_{j=3}^N f_j$, where  $f_j$ results from the summation over the index $i$ for a given value of $j$.  For $j$ close to $\pm n_N$,  corresponding to the Fermi edge, the main contributions to $f_j$ come from few 
values of $i$ that are close enough to $j$ and with the same parity, otherwise the denominator in Eq.~(\ref{F_3expa}) will be large and the contribution negligible. 
To see this, we have calculated $f_j$ from  Eq.~(\ref{F_3expa})  by truncating the sum to $s=1$.  The result is displayed in   Fig. \ref{fig:diff3} as a function of $y=n_j/n_N$,
together with the approximating function $\ln(1+y)-\ln(1-y)$ appearing in Eq.~(\ref{a3finally}). 
Sizeable  differences are seen for $y$ close to $\pm 1$, with the approximation overestimating the
correct result, implying that the logarithmic correction will have negative sign.

We write  $F_3-a_3 N=X+Y+...$, where again $X$ and $Y$ are the contributions to the logarithmic corrections obtained from Eq.~(\ref{F_3expa}) for, respectively, $s=1$ and $s=2$. 
For $s=1$ the summation over the index $i$ in Eq.~(\ref{F_3expa}) reduces to
\begin{equation}
 \sum_{\substack{r=-n_N \\r \ne n_j}}^{n_N}\frac{1}{n_j-r}=h(n_j),
\end{equation}
where the function $h(z)$ is  given by
\begin{equation}
	h(z)=-\psi(1+n_N-z)+\psi(1+n_N+z).
\end{equation}

Taking into account that  $h(z)$ has odd parity,  the contribution to $F_3$ coming from $s=1$ can be
written as an integral via the Euler-Maclaurin formula for Riemann sums
\begin{equation}
	2\sum_{n_j=1}^{n_N-1} \xi^\textrm{th}\left(\frac{n_j}{n_N}\right) h(n_j)+\xi_N h(n_N)\simeq 2n_N  \int_0^1 \xi^\textrm{th}(y)h(n_Ny)dy.
\end{equation}
To isolate the logarithmic correction, we add and subtract the corresponding 
linear-in-$N$ contribution to $F_3$ as given in Eq.~(\ref{a3finally}). This yields
\begin{equation}\label{defX}
	X=2n_N \int_0^1 dy \xi^\textrm{th}(y)  \Big[h(n_Ny)-\ln\left (\frac{1+y}{1-y}\right)\Big] 
\end{equation}

Since the main contributions to the sum in Eq.~(\ref{defX}) come from values of $y$ close to $ 1$, we
can assume that the reduced phase shift is approximately constant,  $\xi^\textrm{th}(y)\simeq \xi^\textrm{th}(1)\equiv \xi_N$. 
The integral in Eq.~(\ref{defX}) can then be calculated analytically yielding
\begin{equation}\label{defXb}
	X=2  \xi_N (-n_N\ln 4 -2 \ln \Gamma(1+n_N)+\ln \Gamma(1+2n_N)).
\end{equation}
Taking into account that $\ln \Gamma(z)\sim z(\ln z-1)$ for $z\gg 1$, we conclude from Eq.~(\ref{defXb})
that $X\simeq -\xi_N \ln N$.

The subleading correction to $F_3$ coming from the contribution with $s=2$  is given by 
\begin{equation}
	Y=\sum_{n_j=-n_N+1}^{n_N}  \frac{1}{2} (\xi^\textrm{th} \left(n_j/n_N\right))^2    \Big[ \sum_{r=-n_N}^{n_N}\frac{1}{(n_j-r)^2}-\frac{\pi^2}{3}\Big],
\end{equation}
where we have subtracted the corresponding linear-in-$N$ correction to $F_3$  in Eq.~(\ref{a3finally}).
By performing the summation over $r$ we find 
\begin{eqnarray}\label{Yagain}
\!\!\!\!\!	Y=\sum_{n_j=-n_N+1}^{n_N} \frac{1}{2} (\xi^\textrm{th} \left(n_j/n_N\right))^2   \Big[ &-&\psi^\prime(1+n_N-n_j)\nonumber \\
	&-&\psi^\prime(1+n_N+n_j) \Big].
\end{eqnarray}
The main contribution to the sum in Eq.~(\ref{Yagain}) comes again from values of $n_j$ close to $\pm n_N$, so we trade  the square of the reduced phase shift with $\xi_N^2$ and approximate  the sum  by an integral
\begin{equation}
Y\simeq \frac{1}{2} \xi_N^2 n_N \int_{-1}^{1} \Big[ -\psi^\prime(1+n_N(1-y))-\psi^\prime(1+n_N(1+y)) \Big]dy.
\end{equation}
The integration over $y$ can then be done analytically yielding 
\begin{equation}
Y\simeq -\xi_N^2  \left (\psi(2n_N+1)-\psi(1) \right),
\end{equation}
showing that  $Y\simeq -\xi_N^2 \ln N$. 

Let us now show that  the contributions with $s>2$ in Eq.~(\ref{F_3expa}) do no produce logarithmic divergences, so that we recover Eq.~(\ref{keyresultF3}). The summation over the index $i$ can be performed analytically so that the contribution of order $s$ becomes
\begin{equation}\label{nolog}
\sum_{j=3}^{N}  \frac{\xi_j^s}{s} \left (\sum_{r=n_N+1}^{+\infty} +\sum_{r=-\infty}^{-n_N-1}\right)
\frac{-1}{(n_j-r)^s}	= \sum_{j=3}^{N}  -\frac{\xi_j^s}{s} R_{n_j},
\end{equation}
where
\begin{equation}\label{R}
	R_{n_j}=\zeta_H(s,1+n_N+n_j)+(-1)^s \zeta_H(s,1+n_N-n_j).
\end{equation}
As for the case $s\leq 2$, we can approximate the reduced phase shifts in Eq. (\ref{nolog}) by $\xi_j\simeq \xi_N (-1)^{(j-1)}$. Since $R_{n_j}=R_{-n_j}(-1)^s$, the 
correction of order $s$ can be written as 
\begin{eqnarray}\label{corr}
&-&\frac{\xi_N^s}{s} \left(\sum_{n_j=1}^{n_N}R_{n_j} +(-1)^s \sum_{n_j=-n_N+1}^{-1} R_{n_j} \right) \nonumber \\
&=&-\frac{\xi_N^s}{s}  \left(2 \sum_{n_j=1}^{n_N}R_{n_j} -(-1)^s R_{-n_N}\right).
\end{eqnarray}
In the limit $n_N\rightarrow +\infty$, the sum over $n_j$ in the second row of Eq.~(\ref{corr}) can be computed analytically so that the correction reduces to
\begin{equation}\label{result}
	(-1)^s\frac{\xi_N^s}{s} (\zeta(s)-2\zeta(s-1)),
\end{equation}
where we have used the identity $R_{-n_N}=\zeta(s)$, which follows from Eq. (\ref{R}). 
 From Eq.~(\ref{result}) one can see that  the correction is finite for  $s>2$, implying that it only affects the constant term in the large-$N$ asymptotics for $F_3$.
  In contrast, for $s=1$ and $s=2$ the correction  diverges, signaling the presence of the logarithmic divergences already discussed.

As a last effort, we  derive the large-$N$ asymptotics of $F_4$ in Eq.~(\ref{central}). Since  $\xi^\textrm{th}(y)$ has odd parity
\begin{eqnarray}\label{F4calc}
	-\sum_{j=3}^N \ln(-\xi_j)&=&-\sum_{\substack{n_j=-n_N+1  \\ n_j \ne 0}}^{n_N}\ln 
	 \xi^\textrm{th}\left(-\frac{n_j}{n_N}\right)\nonumber \\
	 & =&-i\pi n_N-\sum_{\substack{n_j=-n_N+1  \\ n_j \ne 0}}^{n_N}\ln 
	 \xi^\textrm{th}\left(\frac{|n_j|}{n_N}\right).
\end{eqnarray}
The sum in the second row of Eq.~(\ref{F4calc}) can be replaced by an integral, so that
\begin{equation}\label{F4calc1}
	-\sum_{j=3}^N \ln(-\xi_j)\simeq -2n_N \int_0^1 \ln \left (\xi^\textrm{th}(y)\right) dy-i\pi n_N.
\end{equation}
The contributions to $F_4$ containing $\xi_1$ and $\xi_2$ reduce to
\begin{eqnarray}\label{F4calc2}
-\sum_{j=1}^2 \ln  (-\xi_j) &=& -2\ln |\xi_1|, \nonumber \\
	 -\sum_{j=1}^2 \ln (-\xi_j+n_{N})&=&-\ln(|\xi_1|^2+n_N^2), \\
\ln(-\xi_1+\xi_2)&=&\ln (2|\xi_1|)+i\frac{\pi}{2}, \nonumber
\end{eqnarray}
showing that they all scale logarithmically with the system size, but with different coefficients. Notice that the complex contribution in the rhs of the last row makes the determinant of the matrix $C$ purely imaginary, as expected. 

The remaining two sums in Eq.~(\ref{central})  reduce to
\begin{equation}\label{F4calc3}
	-\sum_{j=3}^{N}  \ln(n_{j}+n_{N})+\sum_{j=2}^N \ln(-n_N-n_{j})=i \pi 2 n_N+\ln n_N, 
\end{equation}
whose real part again show a logarithmic divergence for large $N$. Collecting the results in Eq.s (\ref{F4calc1}-\ref{F4calc3})
we  recover Eq.~(\ref{central}), with 
\begin{equation}\label{a4ok}
	a_4=-\int_0^1 \ln \left (\xi^\textrm{th}(y)\right) dy.
\end{equation}

Finally, we can combine Eqs. (\ref{a2ok}), (\ref{a3ok}) and (\ref{a4ok}) to obtain
\begin{equation}\label{sum_ai}
a_2+a_3+a_4=\ln \pi -\int_0^1 \ln (\sin \pi \xi^\textrm{th}(y)) dy.
\end{equation}

Substituting in Eq.~(\ref{sum_ai}) the explicit form for the phase shift given in Eq.~(\ref{xi}) and carrying out the integration leads to
\begin{equation}\label{sum_ai2}
	\sum_r a_r=\ln \pi -\frac{1}{2} \left(2-2 \alpha \arctan\left(\frac{1}{\alpha}\right)+\ln \left(\frac{\alpha^2}{1+\alpha^2}\right)\right).
\end{equation}
By substituting  Eq. (\ref{scalingbv}) as well as Eq.s (\ref{keyresultF2}-\ref{keyresultF4}) back in Eq. (\ref{eqdetV}), we finally obtain the sought expression for the asymptotics  of the determinant of the overlap matrix:
\begin{equation}\label{finalmente}
	\ln \textrm{det}(V) \simeq N \Big(\ln \Big(\frac{g^\prime}{\pi}\Big)-\sum_r a_r \Big) -\Big(\xi_N^2+\frac{1}{2}\Big)\ln N,
\end{equation}
which is the most important result of this section.

\section{Quasi-particle residue }  
	\label{sec:Z}

 Having derived  the large-$N$ asymptotics of the determinants of the norm matrix $S$ and of the overlap matrix $V$,  
  we can now extract the scaling behavior of the quasi-particle residue  
  $Z$,  by substituting Eq. (\ref{detS})  and Eq. (\ref{finalmente}) back  in Eq.~(\ref{Zbis}). This gives
  \begin{eqnarray}\label{Zcalculated}
  Z&\sim &	\left (\frac{|g^\prime|}{\pi}\right)^{2N} \frac{e^{2(\sum_r a_r)N}}{N^{2\xi_N^2+1}} 
  \frac{N}{k_F^{2N} e^{Na}} \nonumber \\
  &\sim & e^{\left [2\ln \left(\frac{|g^\prime|}{\pi k_F}\right)+2 (\sum_r a_r)-a\right]N}N^{-2\xi_N^2}.
  \end{eqnarray}

We next  show that the term inside the square brackets in the second line of Eq.~(\ref{Zcalculated})  actually vanishes, as anticipated in Eq.~(\ref{cond}), so that the quasi-particle residue 
scales as a power of the number  of majority fermions. 
From the explicit expression for the reduced phase shifts in Eq.~(\ref{xi}), we obtain
\begin{equation}
	\sin (\pi \xi^\textrm{th}(y))) =\frac{|\alpha|}{y^2+\alpha^2},
\end{equation}
implying that 
\begin{equation}\label{qui}
	-\int_0^1 \ln (\sin \pi \xi^\textrm{th}(y)) dy= -\ln |\alpha| +\frac{1}{2}\int_0^1 \ln (y^2+\alpha^2) dy.
\end{equation}
A comparison with Eq.~(\ref{res_a_1}) reveals that the second term in the rhs of Eq. (\ref{qui}) is equal to $a/2$. Taking this into account, Eq.~(\ref{sum_ai}) reduces to
\begin{equation}
	\sum_r a_r=\ln \left(\frac{\pi}{|\alpha|}\right) +\frac{a}{2},
\end{equation}
from which the claim immediately follows.
 As a consequence the quasi-particle residue  behaves as 
 \begin{equation} \label{scalingZ}
 	Z =W N^{-\theta},
 \end{equation}  
  where  $W$ is the  prefactor (to be determined numerically below)  and
  the exponent $\theta$ is given by
  \begin{equation}\label{theta}
  	\theta=2\xi_N^2=\frac{2}{\pi^2}\arctan^2\left(-\frac{gm}{2 k_F}\right).
  \end{equation}
  Equation (\ref{scalingZ})  shows that the quasi-particle residue vanishes in the thermodynamic limit,
  which is the signature of the AOC. 
  This result confirms  our recent  findings in Ref.~\cite{OrsoPRA2026}, where the Anderson exponent for the same model was obtained by fitting the numerical data for large $N$   and also by applying boundary conformal field theory arguments \cite{Affleck_JPA1994,AffleckProc1997}.

  It is worth to stress that the formula (\ref{theta}) for the Anderson exponent was previously found to hold for the 1D Fermi polaron with \textsl{repulsive} interactions ~\cite{Castella_1993}, corresponding to $g>0$, where no two-body bound state is present and all  pseudo-momenta $k_i$  are real.  In particular, $\theta$ is independent of the sign of the interaction strength. 
  
\begin{figure}
  	\includegraphics[width=\columnwidth]{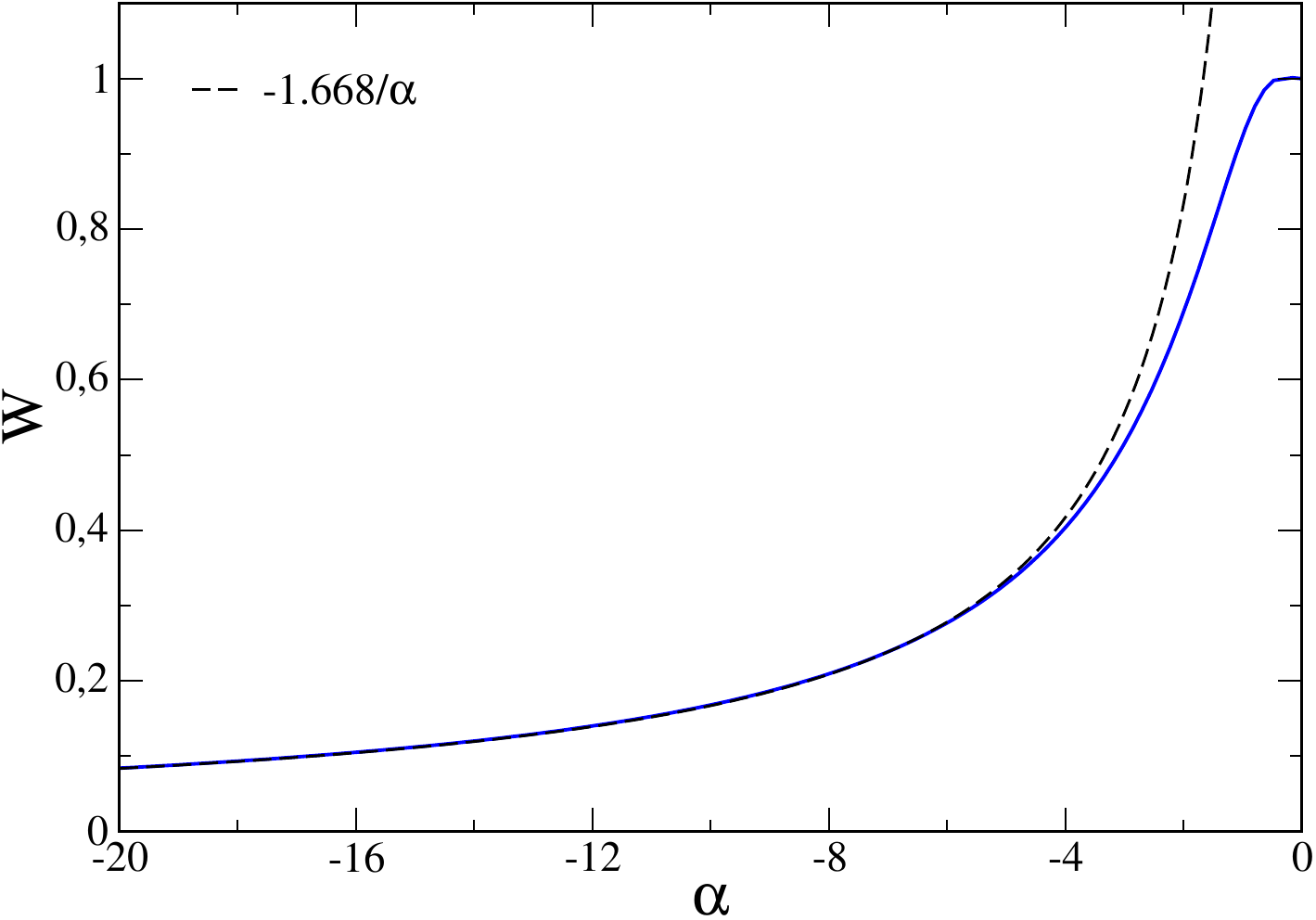}
  	\caption{Prefactor $W$ defined in Eq.~(\ref{scalingZ})  as a function of the interaction parameter. The solid line corresponds to the numerical result, while the dashed line represents  a fit to the data tails.}
  	\label{fig:W}
  \end{figure}

    For completeness, we have extracted  the prefactor $W$ in Eq.~(\ref{scalingZ}) from our  numerical data for large $N$. 
  The result is displayed in Fig.\ref{fig:W} as a function of the interaction parameter (solid line). We see that the prefactor decreases monotonously as the attraction increases and behaves  as $\alpha^{-1}$ for $|\alpha|\gg 1$.  
This scaling behavior can be understood from the asymptotic expressions of  the determinants  of the matrices $S$  and $V$ in the strongly interacting regime $|g|\gg k_F$:
\begin{eqnarray}
	\textrm{det}(S)&= &-\frac{4}{\pi N}(g^\prime)^{(2N)} \frac{k_F}{g^\prime}, \label{eqfirst}\\ 
	\textrm{det}(V)&= & \left(\frac{g^\prime}{\pi}\right)^N b_N \frac{2}{\xi_1}\label{eqsecond} ,
\end{eqnarray} 
 where $b_N$ is a real coefficient that depends only on $N$. Eq. (\ref{eqfirst}) follows directly from Eq.~(\ref{detSS}) by noting that $u_{j}\simeq (g^\prime)^2$
 for $j>2$, according to Eq.~(\ref{uj}). Proving Eq.~(\ref{eqsecond}) is more involved.
 Taking into account that $\xi_1=-\xi_2=-i g^\prime L/(2\pi)$ diverges for infinite attraction,  while $\xi_{j}=(1/2) (-1)^{(j+1)}$ remain finite for all other values of $j$,
we find from Eq.~(\ref{Vbis}) that the matrix $V$ takes the asymptotic form
 \begin{equation}\label{Vfinal}
 V\simeq \frac{g^\prime}{\pi}[\beta-\frac{1}{\xi_1}, \beta+\frac{1}{\xi_1},\cdots],
 \end{equation} 
 where $\beta$ is defined by Eq.~(\ref{beta}) and the dots represent  all columns with index $j$ going from $3$ to  $N$ (such columns  are made by pure numbers). Eq.~(\ref{Vfinal})
 shows that for infinite attraction the first two columns of the matrix $V$ become identical, implying that the determinant must vanish. Computing the determinant of $V$ from 
 Eq.~(\ref{Vfinal}) allows recovering Eq.~(\ref{eqsecond}) with
 \begin{equation}
 	 b_N=\textrm{det}([w,\beta,\cdots]),
 \end{equation}
 where  $w$ is once again an $N$-vector with all components equal to $1$. 
The scaling behavior of the coefficient $b_N$ for large $N$ can be obtained by comparing  Eq. (\ref{eqsecond}) with Eq. (\ref{eqdetV}), yielding  $2 b_N/\xi_1=(1+\beta\cdot v) \textrm{det}(C)$. From Eq. (\ref{scalingbv}) and (\ref{sum_ai2})  we then find
$b_N \sim \pi^N N^{1/4}$, where we used the fact that $\xi_N=1/2$.

  By substituting Eq.s (\ref{eqfirst}) and (\ref{eqsecond}) into Eq.~(\ref{Zbis}), we then find 
 that the quasi-particle residue $Z$ in the strongly attractive regime 
 decays as $1/(\sqrt N \alpha)$, consistently with our numerics.

  A natural extension of this work would be to calculate  the prefactor $W$ analytically,
  for any value of the interaction parameter $\alpha$, by retaining  the next order term in the large-$N$ asymptotics of the two determinants, that is the constant term. 
  This is in principle possible based on the Euler-Maclaurin formula for Riemann sums. We have verified  that   approximating the real phase-shifts in our numerics with the corresponding thermodynamic expression given in Eq.~(\ref{xi}) generally yields a different result for the prefactor $W$.    To recover the correct value, one therefore needs to include  the finite-size corrections to the real phase shifts in the theory. 
  These corrections  can be  calculated from Eq.~(\ref{BAmore1}), by approximating  the
  phase shifts $\delta_j$ in the lhs by their thermodynamic expression in Eq. (\ref{deltaj}),
  yielding
  \begin{equation}
  	\delta_j\simeq \cos^{(-1)}(u_j+\sigma \cos^{(-1)} (u_j)),
  \end{equation}
  with $u_j=- 2n_j/(\alpha N)$. Making use of the Taylor expansion $\cos^{(-1)}(z+\epsilon)\simeq \cos^{(-1)}(z)-\epsilon/(1+z^2)$, holding for small $\epsilon$,
  we obtain
  \begin{equation} \label{corr}
  	\xi_j-\xi^\textrm{th}(x)\simeq -\frac{1}{N} \left(x+\frac{2}{\pi} \xi^\textrm{th}(x)\right)	\frac{\alpha}{\pi}\frac{1}{\alpha^2+x^2},
  \end{equation}
  with $x=n_j/n_N$. Notice that, according to Eq. (\ref{corr}), the finite-size corrections to the phase shifts are positive.
  
  \section{CONCLUSIONS AND OUTLOOK} 
  	\label{sec:conclusion}
  
  In this work we have provided a fully analytical  proof of Anderson's orthogonality catastrophe for the 1D Fermi polaron with attractive interactions, described by the 
   integrable Yang-Gaudin model.
  The Takahashi's representation of the exact
  ground state wave function was used to express the quasi-particle residue $Z$ in terms of the determinants of two matrices, the norm and the overlap matrices. By exploiting the special properties of Cauchy matrices,  we have derived the
  large-$N$ asymptotics of both determinants and show that the expected algebraic decay  $Z=WN^{-\theta}$ emerges from a  subtle cancellation effect. Our exact results confirm the numerical evidence for AOC previously observed in  ~\cite{OrsoPRA2026}, including the
 value of the exponent $\theta$. We have also derived the asymptotic behavior of the 
 prefactor $W$ at strong coupling and contrast it against exact numerics.
  
  The complete excitation spectrum of an impurity interacting with a Fermi sea is accessible experimentally through radio-frequency spectroscopy. In particular the  quasi-particle residue of the polaron can be obtained  through the  measurement of the Rabi oscillation frequency between two spin states of the impurity \cite{Kohstall_2012}. 
 Interacting Fermi gases in 1D configurations have been explored  by using a tight two-dimensional optical lattice to create ensembles of independent wires, which can also be accessed individually \cite{Daniloff:PRL2021}. 
  In these experiments atoms are confined along the wire by a shallow harmonic potential, 
  with the typical  number $N$ of fermions per wire being of the order of one hundred.

 It is also important to stress that the 1D Yang-Gaudin model is realized experimentally by confining ultracold atoms with 3D s-wave interactions in elongated geometries.  In particular the density profiles of spin-imbalanced 1D Fermi gases in a shallow trap have been measured \cite{Liao_2010} and compared to earlier theoretical predictions \cite{Orso:PRL2007a,Hu:PRL2007}, finding good agreement. 
 Simulating the 1D model requires the  conditions $k_B T\ll E_F \ll \hbar \omega_\perp$ and  $E_B \ll \hbar \omega_\perp$ to hold, where $k_B$ is the Boltzmann constant, $T$ is the temperature of the sample,  $E_F=\hbar^2 k_F^2/(2m)$ is the Fermi energy, $\omega_\perp$ is the effective trapping frequency in the transverse direction induced by the 2D optical lattice and we have made explicit the dependence on the reduced Planck constant. The first condition ensures that the motion in the transverse directions is frozen, while the second condition ensures that no significant extra terms in the 1D Hamiltonian are generated 
 by the transverse confinement  through virtual occupation of the excited states \cite{Chevy:PRA2023}. In particular for strong interactions, close to the confinement-induced resonance \cite{Olshanii:PRL1998},  significant deviations from the  predictions based on the Yang-Gaudin model  can occur, as recently observed  with the Chevy's variational ansatz \cite{Barisic:arXiv2025} . Since the variational approach  is unable to see the AOC, an accurate calculation of the spectral weight in the quasi-1D regime remains an interesting open problem for the future.

It would also be worthy to understand whether the present analytical proof of the AOC  can be generalized to other 1D integrable models. 
Another possible direction for future work is to investigate the AOC  for unequal masses of the majority fermions and the impurity, following  ideas  developed in Ref.~\cite{h2f7-dhjh}.

 \begin{acknowledgments}
 	We wish to thank L. Barisic, F. Chevy, E. Gradova and K. Van Houcke for the fruitful collaboration on the Fermi polaron problem in 1D and quasi-1D geometries.  
 	The author acknowledges support from ANR Collaborative project ANR-25-CE47-6400 LowDCertif. 
 	This work was
 	granted access to the HPC resources of TGCC under the
 	allocation AD010513635R3 supplied
 	by GENCI (Grand Equipement National de Calcul Intensif). 
 \end{acknowledgments}
   
\emph{Data availability}. All data of the figures in the manuscript are
available in a Zenodo repository (Ref. \cite{orso_2026_21998838}):
https://doi.org/10.5281/zenodo.21998838.

   \section*{Appendix A. DERIVATION OF NORM AND OVERLAP MATRICES}
 \label{sec:appendixA}  
  
  For vanishing total momentum, $Q=0 $, the many-body wave function $\psi$ of the system depends solely on   the distance coordinates  $y_i=x_i-x_{N+1}$ between each spin-up particle and the impurity, taking the form of a Slater determinant
  \begin{equation}\label{FF}
  	\psi(x_1,\dots,x_{N+1})=  \begin{vmatrix}
  		\varphi_1(y_1)    & \dots & \varphi_N(y_1) \\
  		\vdots & \ddots & \vdots\\
  		\varphi_1(y_N)	 &\dots & \varphi_{N}(y_N)	
  	\end{vmatrix},  
  \end{equation}
  with the single-particle functions $\varphi_i$   being defined as 
  \begin{equation}\label{sp}
  	\varphi_i(y)=(k_i -i g^\prime)e^{i k_i y}- (k_{N+1} -i g^\prime)e^{i k_{N+1} y}, 	
  \end{equation}
  for $i=1,\dots,N$. 
   Setting 	$\langle \varphi_i| \varphi_j \rangle \equiv \int_0^L dy	\varphi_i^*(y)\varphi_i(y) $,
  the square norm of the many-body wave function takes the form 
  \begin{eqnarray}
  	&\langle \psi | \psi\rangle=&  L N!
  	\begin{vmatrix}
  		\langle \varphi_1 | \varphi_1 \rangle    & \dots & 	\langle \varphi_1 | \varphi_N \rangle   \\
  		\vdots & \ddots & \vdots\\
  		\langle \varphi_N | \varphi_1 \rangle 	 &\dots & \langle \varphi_N | \varphi_N \rangle 	
  	\end{vmatrix},
  	\label{identity}
  \end{eqnarray}  
  where the factor $L$ come from the integration over the impurity variable. 
 To reduce overflow issues in numerics as $L$ becomes large,  we write Eq.~(\ref{identity})
  as
  \begin{equation}\label{normpsi}
  	\langle\psi|\psi\rangle=N !\; L^{N +1} \textrm{det}(S)\;,
  \end{equation}  
  where the norm matrix $S$ is defined as
  \begin{eqnarray}
  	&S&=  
  	\begin{vmatrix}
  		\frac{1}{L} \langle \varphi_1 | \varphi_1 \rangle    & \dots & 	\frac{1}{L} \langle \varphi_1 | \varphi_N \rangle   \\
  	  \vdots & \ddots & \vdots\\
  		\frac{1}{L} \langle \varphi_N | \varphi_1 \rangle 	 &\dots & \frac{1}{L} \langle \varphi_N | \varphi_N \rangle 	
  	\end{vmatrix}.
  	\label{identity2}
  \end{eqnarray}

The 1D overlaps in Eq.(\ref{identity2}) can be evaluated analytically from Eq.~(\ref{sp}) yielding
\begin{eqnarray}\label{eqlong}
	\frac{1}{L}\langle \varphi_i | \varphi_j \rangle  &=&(k_i^* +i g^\prime)
	(k_j -i g^\prime)  \tilde R_L(k_j-k_i^*) \nonumber \\
	&-& (k_i^* +i g^\prime)
	(k_{N+1} -i g^\prime) \tilde R_L(k_{N+1}-k_i^*) \\
	&-& (k_{N+1}^* +i g^\prime)
	(k_j -i g^\prime) \tilde R_L(k_j-k_{N+1}^*) \nonumber \\
	&+& (k_{N+1}^* +i g^\prime)
	(k_{N+1} -i g^\prime) \tilde R_L(k_{N+1}-k_{N+1}^*) \nonumber, 
\end{eqnarray}
with the function $\tilde R_L(p)$ being defined as
\begin{equation}\label{defR}
	\tilde R_L(p)=\frac{1}{L}\int_0^L e^{i p y}dy=
	\begin{cases} 
		\frac{(e^{i pL}-1)}{ipL} & \textrm{if}\; p\neq 0 \\
		1 & \textrm{if}\; p=0 \; .\\
	\end{cases}	
\end{equation}  
For $N\ge 3$ the quasi-momentum $k_{N+1}$ in Eq.~(\ref{eqlong}) is by contruction real, that is $k_{N+1}^*=k_{N+1}$.  The first contribution in the rhs of Eq.~(\ref{eqlong}) can be evaluated by rewriting the Bethe-ansatz equation (\ref{BA1})
as $(k_i-i g^\prime)e^{ik_i L}=k_i+ig^\prime$ and taking the complex conjugate.  From Eq.~(\ref{defR}) we then find
\begin{equation}\label{III}
(k_i^* +i g^\prime)
(k_j -i g^\prime)  \tilde R_L(k_j-k_i^*)=	\begin{cases} 
	-\frac{2g^\prime}{L}& \textrm{if}\; k_i^*\neq k_j \\
	k_j^2+g^{\prime 2} & \textrm{if}\;  k_i^*= k_j \; .\\
\end{cases}	
\end{equation}
By the same token, one can show that the second and the third contributions in the rhs of 
Eq.~(\ref{eqlong}) both reduce to $2g^\prime/L$, while the last contribution reduces to 
$k_{N+1}^2+g^{\prime 2}$. Taking into account Eq.~(\ref{III}) and Eq.~(\ref{uj}),  we obtain
\begin{equation}
	S_{ij}=\begin{cases} 
		u_{N+1}& \textrm{if}\; k_i^*\neq k_j \\
		u_{N+1}+u_j & \textrm{if}\;  k_i^*= k_j \; ,\\
	\end{cases}
\end{equation}
which is consistent with the definition of the norm matrix  given in Eq.(\ref{def_S}).
   
In the absence of interaction, the  many-body wave function 
is the product of the plain wave Slater determinants of the two  spin components. 
\begin{equation}\label{psiNI}
	\psi_{NI}(x_1,..,x_{N+1})=	\begin{vmatrix}
		e^{i q_2 x_1} & \dots & e^{i q_{N+1}x_1} \\
		\vdots & \ddots & \vdots\\
		e^{i q_2 x_{N}}  & \dots & e^{i q_{N+1}x_{N}} \\	
	\end{vmatrix}
	e^{i q_1 x_{N +1}} \; , 
\end{equation}
where the fermion quasi-momenta $q_j$ satisfy the condition 
$e^{i q_j L}=1$, yielding $q_j=e^{i 2\pi n_j/L}$, where $n_j$ is an integer.
The ground state  corresponds to the choice 
\begin{equation}
\{n_j\}_{j=1 \cdots N+1}=\{0,0, \cdots,\frac{(N-1)}{2},-\frac{(N-1)}{2}\},
\end{equation}
ensuring that the fermion quasi-momenta $\{k_j\}$ reduce to $\{q_j\}$ for vanishing interaction, $g^\prime=0$.

Written in terms of the distance variables, the wave function in Eq.~(\ref{psiNI})  takes the form 
\begin{equation}
	\psi_{NI}(x_1,..,x_{N+1})=\begin{vmatrix}
		e^{i q_2 y_1} & \dots & e^{i q_{N+1}y_1} \\
		\vdots & \ddots & \vdots\\
		e^{i q_2 y_{N}}  & \dots & e^{i q_{N+1}y_{N}} \\	
	\end{vmatrix} \; ,
\end{equation}
from which it easily follows that
\begin{equation}\label{normpsiNI}
	\langle\psi_{NI}|\psi_{NI}\rangle=N !\; L^{N +1}\; .    
\end{equation}

The overlap between the two ground states can be evaluated
from Eq.~(\ref{FF}) and Eq.~(\ref{psiNI}) as
\begin{equation}\label{over}
	\langle\psi_{NI}|\psi\rangle=N!\; L^{N+1} \textrm{det}(V)\; ,    
\end{equation}
where the entries of the overlap matrix $V$  are given by
\begin{equation}\label{Vapp}
	V_{ij}=\frac{1}{L} \int_0^L e^{-i q_{i+1}y} \varphi_j(y) dy. 
\end{equation}
Substituting Eq.~(\ref{sp}) in Eq.~(\ref{Vapp}) and carrying out the integration 
leads to
\begin{equation}\label{Vapp2}
	V_{ij}=(k_j-i g^\prime) \tilde R_L(k_j-q_{i+1})-(k_{N+1}-i g^\prime) \tilde R_L(k_{N+1}-q_{i+1}).
\end{equation}
Making use of Eq.~(\ref{defR}) and recalling that $e^{-i q_{i+1}L}=1$, we can 
recast Eq.~(\ref{Vapp2}) as
\begin{equation}
	V_{ij}=\frac{2g^\prime}{(k_j-q_{i+1})L}-\frac{2g^\prime}{(k_{N+1}-q_{i+1})L},
\end{equation}
which coincides with Eq.~(\ref{V}). Finally, substituting Eq.s (\ref{normpsi}), (\ref{normpsiNI}) and (\ref{over}) in Eq.~(\ref{Z}) allows recovering Eq.~(\ref{Zbis}). 

   \section*{Appendix B. CAUCHY MATRICES}
   	\label{sec:appendix}
   
A square matrix $C$ is said to be a Cauchy matrix if its elements are of the form
\begin{equation}
	C_{ij}=\frac{1}{x_i+y_j},
\end{equation}
where $x_i$ and $y_j$ are numbers  satisfying $x_i+y_j \ne 0$ for every pair $i,j$ with $i,j=1 \dots N$. 

The determinant of a Cauchy matrix $C$ can  be written as
\begin{equation}\label{cauchydet}
\det(C)=\frac{\prod_{i<j} (x_i-x_j)(y_i-y_j)}{\prod_{i,j} (x_i+y_j)},
\end{equation}
showing that the determinant vanishes if  $x_i=x_j$ or $y_i =y_j$ for some values
of $i \ne  j$, as the matrix $C$ will have two equal rows or two equal columns.

If the Cauchy matrix $C$ is invertible, the entries of the inverse matrix $C^{-1}$ are given by
\begin{equation} \label{cauchyinv}
(C^{-1})_{ij}=(x_j+y_i) \frac{\prod_{\ell \ne i} (x_j+y_\ell) \prod_{k \ne j } (y_i+x_k)}{\prod_{\ell \ne j} (x_j-x_\ell) \prod_{k \ne i} (y_i-y_k)},
\end{equation}
from which one can shown that
\begin{equation}\label{chauchyvj}
	\sum_{i} (C^{-1})_{ij}=\frac{\prod_\ell (x_j+y_\ell)} {\prod_{\ell \ne j} (x_j-x_\ell)  }.
\end{equation}


\bibliography{main}

\end{document}